\documentclass[prd,twocolumn,amsmath,amssymb,showpacs,preprintnumbers,nofootinbib]{revtex4}
\usepackage{graphicx,dcolumn}

\newcolumntype{.}{D{.}{.}{8}}

\newcommand{\lnx}{\ln x}
\newcommand{\qvec}{|\vec{q}\,|}
\newcommand{\ep}{\epsilon}
\newcommand{\order}[1]{\mathcal{O}\left( #1 \right)}
\newcommand{\UplusL}{}

\begin{document}

\title{\boldmath Helicity fractions of $W$ bosons from top quark decays
  at NNLO in QCD}

\author{Andrzej Czarnecki}
\email{andrzejc@ualberta.ca}
\affiliation{Department of Physics, University of Alberta, Edmonton,
  Alberta T6G 2G7, Canada}
\affiliation{CERN Theory Division, CH-1211 Geneva 23, Switzerland}
\author{J{\"u}rgen G. K{\"o}rner}
\email{koerner@thep.physik.uni-mainz.de}
\affiliation{Institut f{\"u}r Physik, Universit{\"a}t Mainz, 55099
  Mainz, Germany}
\author{Jan H. Piclum}
\email{jpiclum@phys.ualberta.ca}
\affiliation{Department of Physics, University of Alberta, Edmonton,
  Alberta T6G 2G7, Canada}

\date{\today}

\begin{abstract}
  Decay rates of unpolarized top quarks into longitudinally and
  transversally polarized $W$ bosons are calculated to second order in
  the strong coupling constant $\alpha_s$. Including the finite
  bottom quark mass and electroweak effects, the Standard Model
  predictions for the $W$ boson helicity fractions are ${\cal F}_{L}=
  0.687(5)$, ${\cal F}_{+}= 0.0017(1)$, and ${\cal F}_{-} = 0.311(5)$.
\end{abstract}


\preprint{ALBERTA-THY-05-10}
\preprint{CERN-PH-TH-2010-098}
\preprint{MZ-TH/10-13}

\maketitle


There has been a continuing interest in the measurement of the helicity
fractions of the $W$ boson from top quark decays from the CDF
collaboration~\cite{Affolder:1999mp,Acosta:2004mb,Abulencia:2005xf,%
Abulencia:2006iy,Abulencia:2006ei,Aaltonen:2008ei,Aaltonen:2010ha} and
from the D0 collaboration~\cite{Abazov:2005fk,Abazov:2004ym,%
Abazov:2006hb,Abazov:2007ve} at the Tevatron at Fermilab.

In the Standard Model (SM) the top quark decays predominantly into a
$W^{+}$ boson and a bottom quark.  Interesting observables, independent
of the production rate that is difficult to predict precisely for a
hadron collider, are the fractions of the three possible $W$ helicities:
${\cal F}_{L}$ (longitudinal), ${\cal F}_{+}$ (transverse-plus) and
${\cal F}_{-}$ (transverse-minus). In the  leading order (LO) in the
strong coupling constant $\alpha_s$ (that is, without any gluon
corrections), and in the limit of a massless bottom quark one
has~\cite{Kane:1991bg}
\begin{equation}
  \label{fractions}
        {\cal F}_{L}:{\cal F}_{+}:{\cal F}_{-}=\frac{1}{1+2x^{2}}:0
        :\frac{2x^{2}}{1+2x^{2}} \,\,  ,
\end{equation}
with ${\cal F}_{L}+{\cal F}_{+}+{\cal F}_{-}=1$ and $x\equiv m_{W}/m_{t}$.
Using $m_{W}=80.401(43)$~GeV~\cite{Abazov:2009cp} and
$m_{t}=172.8(1.3)$~GeV~\cite{CDF2010} we get
$x^{2}=m_{W}^{2}/m_{t}^{2}=0.216(3)$ and
${\cal F}_{L}:{\cal  F}_{+}:{\cal F}_{-}\simeq 0.7:0:0.3$.  

The leading order decay $t\to bW$ is a two-body process.  With the $V-A$
interaction, a massless $b$ quark is left-handed, thus the $W$ can only
be left-handed or longitudinal due to angular momentum conservation. One
therefore has ${\cal F}_{+}=0$ provided no gluons are emitted.  

The above LO  predictions are only marginally changed by the bottom
mass. For a pole mass of $m_{b}=4.8$~GeV one finds that the total rate
$\Gamma_{\UplusL}$ decreases by about a quarter per cent compared to
the massless $b$ limit.  The helicity fraction  ${\cal F}_{L}$
slightly decreases while  ${\cal F}_{-}$ increases, by about one per
mil.  The  leakage into the transverse-plus
fraction ${\cal F}_{+}$ is less than half per mil.
Radiative corrections are a more important source of the
transverse-plus rate. However, as we shall see, when NLO and NNLO
gluon radiation is included,  
${\cal F}_{+}$ still does not exceed two per mil.  Since only hard
gluon emission can influence the helicity fractions, this smallness is
a reliable prediction of the Standard Model.

For this reason,
the transverse-plus fraction ${\cal F}_{+}$ is a sensitive probe of New
Physics effects such as a right chiral admixture to
the SM current. The left and right chiral contributions do not interfere
for $m_{b}=0$ leading to a quadratic dependence on the admixture
parameter. The contribution of the right chiral contribution can be
obtained from Eq.~(\ref{fractions}) by exchanging 
${\cal F}_{+}\leftrightarrow{\cal F}_{-}$ whereas ${\cal F}_{L}$ remains
unchanged. We mention that there are some indirect model dependent
constraints on a possible right chiral admixture to the SM current from
measurements of $b \to s + \gamma$ decays~\cite{Fujikawa:1993zu,%
Cho:1993zb,AguilarSaavedra:2007rs,Grzadkowski:2008mf}.

Let us summarize the theoretical prediction for the helicity
fractions. In addition to the $\order{\alpha_{s}^{2}}$ effects computed
in this paper, we include the lower-order contribution (Born +
$\order{\alpha_{s}}$)~\cite{Fischer:1998gsa,Fischer:2000kx,Fischer:2001gp},
the leading electroweak corrections~\cite{Do:2002ky} and, for
${\mathcal{F}}_+$,  the $m_b$ effect. The errors resulting from
uncertainties in  $m_{t,b,W}$ and $\alpha_{s}$ and an estimate of the
higher-order effects are added in quadrature. We find
\begin{eqnarray}
  {\cal F}_{L}&=& 0.687(5)\,, \nonumber \\
  {\cal F}_{+}&=& 0.0017(1)\,, \nonumber \\
  {\cal F}_{-}&=& 0.311(5)\,.
\end{eqnarray}
The relative errors for ${\cal F}_{L}$ and ${\cal F}_{-}$ are small (of
$\order{1\%}$) and, for the largest part, result from the experimental
error on the top mass. The error for ${\cal F}_{+}$ arises from
uncertainty in
$\alpha_s$ and, to a lesser degree, in $m_b$.  Its absolute value is small
but the relative error is large due to the fact that ${\cal F}_{+}$
vanishes at LO for $m_b = 0$.

Various methods have been used by the CDF and D0 collaborations to
experimentally extract the helicity fractions from the top quark decay
data (see the recent review~\cite{Wagner:2010wd}). While previous
analyses have performed two fits keeping one of the helicity fractions
at its SM value more recent analyses measure the fractions ${\cal F}_{L}$
and ${\cal F}_{+}$ simultaneously in a two-dimensional
fit~\cite{Aaltonen:2010ha,Abazov:2007ve}. Using such a model independent
analysis the CDF collaboration quotes values of
${\cal F}_{L}=0.88 \pm 0.11({\rm stat})\pm 0.06({\rm syst})$ and
${\cal F}_{+}=-0.15 \pm 0.07({\rm stat}) \pm 0.06({\rm syst})$~\cite{Aaltonen:2010ha}.
In a similar analysis the D0 collaboration obtains
${\cal F}_{L}=0.425 \pm0.166({\rm stat})\pm0.102({\rm syst})$ and
${\cal F}_{+}=0.119 \pm 0.090 ({\rm stat})\pm0.053({\rm syst})$~\cite{Abazov:2007ve}.
Both measurements are consistent with the SM predictions.

The experimental errors on the helicity fraction measurements are still
rather large but will be much reduced when larger data samples become
available in the future from the Tevatron and from the LHC. 
Optimistically the measurement errors can eventually be
reduced to below one per cent. For example, an early Monte Carlo
(MC) study quotes measurement uncertainties of
$\Delta{\cal F}_{L}=0.007$ and $\Delta{\cal F}_{+}=0.003$ for an
integrated luminosity of 100 ${\rm fb}^{-1}$ at Tevatron II
energies~\cite{Amidei:1996dt}. The corresponding event rates can easily
be reached at the LHC within one year. A more recent MC study based on
10 ${\rm fb}^{-1}$ at the LHC quotes measurement uncertainties of
$\Delta{\cal F}_{L}=0.019$, $\Delta{\cal F}_{-}=0.018$ and
$\Delta{\cal F}_{+}=0.0021$~\cite{AguilarSaavedra:2007rs}.

The improvements in the accuracy of the experimental measurements
have to be matched by corresponding advances in the theoretical
sector. The NLO $\order{\alpha_{s}}$ corrections to the helicity
fractions were calculated
in~\cite{Fischer:1998gsa,Fischer:2000kx,Fischer:2001gp}. 
They lower  ${\cal F}_{L}$  and
increase  ${\cal F}_{-}$ by about one and two per cent, respectively,
relative to their LO values. 
At NLO there is now a small
contribution to the transverse-plus fraction ${\cal F}_{+}$ of
$0.001$. The corresponding NLO electroweak and finite width corrections
were determined in~\cite{Do:2002ky}. They are smaller than the strong
corrections and tend to cancel each other for both ${\cal F}_{L}$ and
${\cal F}_{-}$.

It is desirable to improve the accuracy of the theoretical predictions
and to check the convergence of the perturbative series by computing the
helicity fractions at NNLO. A first step in this direction was taken
in~\cite{Blokland:2004ye,Blokland:2005vq}~\footnote{An approximate value
  for the total rate was found in Ref.~\cite{Chetyrkin:1999ju}.}
where the NNLO corrections to the total rate were found, exploiting the
smallness of $x=m_{W}/m_{t}$. A series in powers and logarithms of $x$
was obtained and found to converge rapidly, so that its first few terms
suffice. The aim of this paper is to use similar techniques to
calculate the NNLO strong corrections to the three helicity fractions.

We first determine the rate $\Gamma_L$ of the top decay with
longitudinally polarized $W$, replacing the full sum over $W$
polarizations by a projector described below. The previous knowledge of
the total rate is used to calculate the transverse rate
$\Gamma_{T}=\Gamma_{+}+\Gamma_{-}$ from the difference
$\Gamma_{T}= \Gamma_{\UplusL}-\Gamma_{L}$. We use another projector to
find the difference $\Gamma_{+}-\Gamma_{-}$. Finally, the helicity
fractions ${\cal F}_{i}=\Gamma_{i}/\Gamma_{\UplusL}$ are determined.


\begin{figure}[t]
  \includegraphics[width=\columnwidth]{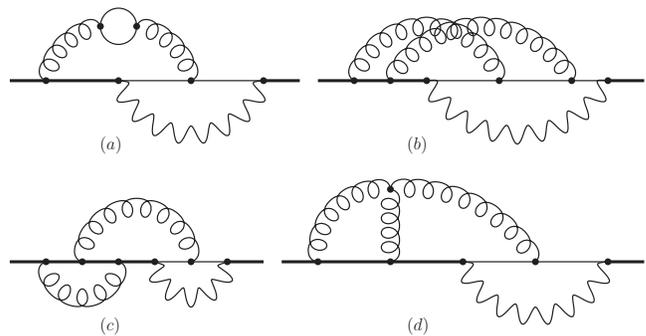}
  \caption{\label{fig::dias} Sample three-loop diagrams. Thick and thin
    lines denote top and bottom quarks, respectively. Wavy lines denote
    $W$ bosons and curly lines denote gluons. In the closed fermion loop
    all quark flavors have to be considered.}
\end{figure}

Our calculation follows the approach outlined in
Ref.~\cite{Blokland:2005vq}. Using the optical theorem we compute
the decay width from the imaginary part of top-quark self-energy
diagrams,
\begin{equation}
  \Gamma = \frac{1}{m_t}\, {\rm Im} \left( \Sigma \right) \,,
  \label{eq::optical}
\end{equation}
where $\Sigma$ denotes the one-particle irreducible self-energy
diagrams. Sample diagrams are shown in Fig.~\ref{fig::dias}. The unitary
gauge is used for the $W$ boson so that diagrams with Goldstone bosons
are not needed. However, the $R_\xi$ gauge is used for the gluons with
an arbitrary gauge parameter. The gauge-parameter dependence cancels in
the final result.

Since we set the mass of the bottom quark to zero, the integrals contain
two scales, $m_t$ and $m_W$. To reduce these integrals to single scale
integrals, we use the method of expansion by regions (see, e.g.,
Ref.~\cite{Smirnov:2002pj}). In the present case, there are two regions
to be considered. In the so-called hard region, the loop momenta are of
the order of $m_t$, while they are of order $m_W$ in the so-called soft
region. The integrals become scaleless and vanish if a gluon momentum is
soft. Thus, we are left with two contributions to each integral: one
where all momenta are hard and one where only the $W$-boson momentum is
soft. For each contribution we construct appropriate expansions in the
corresponding small quantities. The remaining single scale integrals are
further reduced to so-called master integrals using Laporta's
algorithm~\cite{Laporta:1996mq,Laporta:2001dd}.

Compared to the NNLO calculation of the total width, to get the partial
rates with various $W$ polarizations requires  replacing the total rate
projector~\footnote{In the unitary gauge, the $W$ boson propagator reads
  $i\, \mathbb{P}^{\mu\nu}_{\UplusL}/(q^2 - m_W^2)$.}
\begin{equation}
  \mathbb{P}^{\mu\nu}_{\UplusL} = - g^{\mu\nu} + \frac{q^\mu
    q^\nu}{m_W^2} \,,
  \label{eq::proj_unpol}
\end{equation} 
by the longitudinal projector $\mathbb{P}^{\mu\nu}_L$ or the
transverse-plus/minus projectors $\mathbb{P}^{\mu\nu}_{\pm}$. The
longitudinal projector reads \cite{Fischer:2000kx}
\begin{equation}
  \mathbb{P}^{\mu\nu}_L = \frac{(m_W^2\, p^\mu
    - p\cdot q\, q^\mu)(m_W^2\, p^\nu - p\cdot q\, q^\nu)}{m_W^2 m_t^2\,
    \qvec^2}\, ,
  \label{l}
\end{equation}
where $p$ is the top quark momentum and $q = (q_0;0,0,\qvec)$ is the
momentum of the $W$ boson which propagates in the $z$-direction. The
transverse projectors can be obtained with the help of the
forward-backward projector \cite{Fischer:2000kx}
\begin{equation}
  \mathbb{P}^{\mu\nu}_F = -\frac{1}{m_t \qvec}\,
  i\varepsilon^{\mu\nu\sigma\rho} p_\sigma q_\rho \,.
  \label{fb}
\end{equation}
One has $\mathbb{P}^{\mu\nu}_{\pm}= \left(\mathbb{P}^{\mu\nu} -
\mathbb{P}^{\mu\nu}_L \pm \mathbb{P}^{\mu\nu}_F \right)/2$.

Partial helicity rates involve two technical challenges absent in the
total rate calculation. First, there is an additional propagator-like
structure $1/\qvec^n$, $n\in \{1,2\}$ in Eqs.~(\ref{l}) and (\ref{fb}),
and second, we have to deal with the presence of $\gamma_{5}$-odd traces
in dimensional regularization. Our approach to both issues is outlined
below.

In the hard region, we express $\qvec^2$ through the propagator factor
$N=(p+q)^{2}-m_{t}^{2}=2pq+q^{2}$ as follows~\footnote{$p=(m_{t};0,0,0)$
  in the rest frame of the top quark such that $p\cdot q=m_{t}q_{0}$.}:
\begin{eqnarray}
  \qvec^2 &=& q_0^2 - m_W^2 = \frac{(2 p\cdot q)^2}{4m_t^2} -
  m_W^2  \nonumber \\
  &=& \frac{1}{4 m_t^2} \left[ N^2 - 2
    m_W^2 N + m_W^4 - 4 m_t^2 m_W^2
    \right] \,.
  \label{eq::oncut}
\end{eqnarray}
In Eq.~(\ref{eq::oncut}) we use the fact that we are only interested in
the imaginary part and that $q^2 = m_W^2$ on the cut. Now we can
construct the desired expansions in $m_W/m_t$ as
\begin{eqnarray}
  \frac{1}{\qvec} &=& \frac{2 m_t}{N}
  \sum_{i=0}^\infty \binom{2i}{i}\!\! \left( \frac{2 m_W^2 N 
    - m_W^4 + 4 m_t^2 m_W^2}{4\,N^{2}}
  \right)^i \,, \nonumber \\
  \label{eq::exp_fba}
  \frac{1}{\qvec^2} &=& \frac{4 m_t^2}{N^{2}}
  \sum_{i=0}^\infty \left( \frac{2 m_W^2 N
    - m_W^4 + 4 m_t^2 m_W^2}{N^2} \right)^i \,,
  \label{eq::exp_long}
\end{eqnarray}
which we truncate at some order. Thus, the additional propagator-like
structure from the projector is transformed into a scalar on-shell
propagator with momentum $p+q$ and mass $m_t$, raised to arbitrary,
integer powers. For the calculation of the polarized decays we need,
next to the master integrals of Refs.~\cite{Blokland:2005vq,Pak:2006xf},
twelve additional three-loop master integrals.

In the soft region, we cannot perform an expansion of $\qvec$, since
$\qvec^2 = q_0^2 - m_W^2$ and $q_0$ is of order $m_W$ in the soft
region. However, in this region the $W$-boson loop
factorizes. Therefore, we only have to replace the usual one-loop
tadpole integrals with integrals of the type
\begin{equation}
  \int \frac{{\rm d}^dq}{(q^2 - m_W^2)\, (q_0^2 - m_W^2)^n} \,,
  \label{eq::softint}
\end{equation}
with $n \in \{1/2,1\}$. $d = 4 - 2\ep$ is the number of
dimensions. Integrals of this type can be easily evaluated by performing
the integrations over the time-like and space-like momentum components
separately.

For traces with an odd power of $\gamma_5$, we use the prescription of
Ref.~\cite{Larin:1993tq} and replace
\begin{equation}
  \gamma_\mu\gamma_5 \to \frac{i}{3!} \varepsilon_{\mu\alpha\beta\delta}
  \gamma^\alpha \gamma^\beta \gamma^\delta \,.
  \label{eq::gamma5}
\end{equation}
The $\varepsilon$ tensor is stripped off and absorbed into the
projector. As a consequence the renormalization constant of the
axial-vector current at the requisite order becomes
\begin{equation}
  Z_A = 1 + \left( \frac{\alpha_s}{\pi} \right)^2 \left( \frac{11}{24}
  C_F C_A - \frac{1}{6} C_F T_F n_f \right) \frac{1}{\ep} \,,
  \label{eq::za}
\end{equation}
where $C_F = (N_c^2-1)/(2N_c)$ and $C_A = N_c$ are the Casimir operators
of the fundamental and adjoint representation of SU$(N_c)$,
respectively. For QCD we have $N_c = 3$ and $T_F = 1/2$. $n_f$ denotes
the number of quark flavors. Additionally, we have to include the finite
renormalization constant
\begin{eqnarray}
  Z_5 &=& 1 - \frac{\alpha_s}{\pi} C_F + \left( \frac{\alpha_s}{\pi}
  \right)^2 \bigg( \frac{11}{8} C_F^2 - \frac{107}{144} C_F C_A
  \nonumber \\
  &&+\frac{1}{36} C_F T_F n_f \bigg) 
  \label{eq::z5}
\end{eqnarray}
to restore the anti-commutativity of $\gamma_5$. Both renormalization
constants were determined at the three-loop level in
Ref.~\cite{Larin:1991tj}.

A NLO check of the new methods used in this paper is afforded by
comparing with the expanded form of the known NLO closed form results
given in~\cite{Fischer:1998gsa,Fischer:2000kx,Fischer:2001gp}. We found
agreement up to $\order{x^{16}}$.


We present our results in terms of the reduced helicity rates
$\hat{\Gamma}_{i}$ where
\begin{equation}
  \Gamma_i = \frac{G_F m_t^3 |V_{tb}|^2}{8 \sqrt{2} \pi} 
  \hat{\Gamma}_i
\end{equation}
with $i \in \{L,+,-\}$. $G_F$ is Fermi's constant and $V_{tb}$ is the
element of the Cabibbo-Kobayashi-Maskawa matrix which governs
transitions between bottom and top quarks.

The analytical results of our calculation are too long to be presented
here. Instead we present their numerical values. In
Table~\ref{tab::numerics} we show successive terms in the power series
expansion (in terms of $[x^{n}]:=(x^{n},x^{n}\lnx)$) of the NNLO
correction to the reduced helicity rates up to terms of order
$\order{[x^{10}]}$. Noteworthy is the fact that there are also odd
powers in the expansion in $[x]$. These terms appear in the expansion of
the parity-odd helicity structure function $\Gamma_{F}$~\footnote{This
  follows the pattern in unpolarized and polarized top quark decays
  where the expansion of the five parity-even structure functions have
  $n$=even whereas the expansion of the five parity-odd structure
  functions have $n$=even/odd~\cite{Fischer:2000kx}.} and thereby in
$\Gamma_{\pm}$ (see Table~\ref{tab::numerics}). The odd powers of $x$
stem solely from the soft region of $\Gamma_F$. The leading $x^{0}$
contributions of $\hat{\Gamma}_{L}$ and $\hat{\Gamma}_{\UplusL}$ are
equal to each other. This is a consequence of the Goldstone boson
equivalence theorem. Between the $[x^{4}]$ and $[x^{6}]$ terms the power
series expansion is somewhat erratic for $\hat{\Gamma}_{+}$,
$\hat{\Gamma}_{-}$ and $\hat{\Gamma}_{\UplusL}$. However, expanding up
to $\order{[x^{10}]}$ Table~\ref{tab::numerics} shows that one has
sufficient numerical stability and precision for all three helicity
rates and their sum. The contribution of the $\order{[x^{10}]}$ term
amounts to about $0.01 \%,\,1\%,\,0.06\%$ and $0.03\%$ of the total for
$\hat{\Gamma}_{L},\hat{\Gamma}_{+}, \hat{\Gamma}_{-}$ and
$\hat{\Gamma}_{\UplusL}$, respectively. The convergence is slowest for
$\hat{\Gamma}_{+}$. But then $\hat{\Gamma}_{+}$ is numerically very
small.

\begin{table}[h]
  \caption{\label{tab::numerics}
    Numerical values for $\order{[x^{n}]:=(x^{n},x^{n}\ln x)}$ terms in
    the $x$--expansion of the NNLO corrections to the reduced partial
    and total helicity rates ${\hat \Gamma}_{i}$}.
  \begin{ruledtabular}
    \begin{tabular}{l....}
      \phantom{$\hat{\hat{I}_3}$} &
      \multicolumn{1}{c}{$\hat{\Gamma}_{L}$} &
      \multicolumn{1}{c}{$\hat{\Gamma}_{+}$} &
      \multicolumn{1}{c}{$\hat{\Gamma}_{-}$} &
      \multicolumn{1}{c}{$\hat{\Gamma}_{\UplusL}$}  \\ 
      \hline \\[0.1ex]
      $[x^{0}]$& -1.958\cdot 10^{-2}& 0& 0& -1.958\cdot 10^{-2} \\[0.5ex]
      $[x^{2}]$& 4.737\cdot 10^{-3}&3.860\cdot 10^{-4} &-3.861\cdot 10^{-3}
      & 1.262\cdot 10^{-3} \\[0.5ex]
      $[x^{4}]$& 6.710\cdot 10^{-4}& 1.351\cdot 10^{-4}& 
      -9.917\cdot 10^{-4}& -1.856\cdot 10^{-4} \\[0.5ex]
      $[x^{5}]$& 0& -5.339\cdot 10^{-4}& 5.339\cdot 10^{-4}& 0 \\[0.5ex]
      $[x^{6}]$&-1.467\cdot 10^{-4} &1.186\cdot 10^{-4} & 
      4.878\cdot 10^{-4}& 4.597\cdot 10^{-4} \\[0.5ex]
      $[x^{7}]$& 0& 7.696\cdot 10^{-5}& -7.696\cdot 10^{-5}& 0 \\[0.5ex]
      $[x^{8}]$& -1.702\cdot 10^{-5}& -2.333\cdot 10^{-5}& 
      -1.723\cdot 10^{-5}& -5.758\cdot 10^{-5} \\[0.5ex]
      $[x^{9}]$& 0& 3.408\cdot 10^{-6}& -3.408\cdot 10^{-6}& 0 \\[0.5ex]
      $[x^{10}]$& -1.274 \cdot 10^{-6}& -1.844\cdot 10^{-6}&
      -2.176\cdot 10^{-6} & -5.294\cdot 10^{-6} \\[0.5ex]
      \hline \\[0.1ex]
      $\Sigma$ &-1.434\cdot 10^{-2}&1.610\cdot 10^{-4}&-3.931\cdot 10^{-3}&
      -1.811\cdot 10^{-2}
    \end{tabular} 
  \end{ruledtabular}
\end{table}

In order to present our numerical results on the helicity fractions we
define helicity fractions up to $\order{n}$ by writing ($n=0,1,2$ denote
the contributions up to LO, NLO and NNLO, respectively)
\begin{equation}
  \label{helfracdef}
        {\cal F}_{i}^{(n)} 
        = \frac{\sum_{j=0}^{n} \Gamma_{i}^{(j)}}
        {\sum_{j=0}^{n} \Gamma_{\UplusL}^{(j)}},
\end{equation}
where $i=L,+,-$. We further define the increments 
$\Delta{\cal F}_{i}^{(n)}= {\cal F}_{i}^{(n)}-{\cal F}_{i}^{(n-1)} $ and
the relative increments
$\delta{\cal F}_{i}^{(n)}=\Delta{\cal F}_{i}^{(n)}/{\cal F}_{i}^{(0)}$.
We present our numerical results in the form
${\cal F}_{i}= {\cal F}_{i}^{(0)} + \Delta{\cal F}_{i}^{(1)}
+\Delta{\cal F}_{i}^{(2)}$, and also, if ${\cal F}_{i}^{(0)} \neq 0$, as
${\cal F}_{i}= {\cal F}_{i}^{(0)} (1 + \delta{\cal F}_{i}^{(1)}
+\delta{\cal F}_{i}^{(2)}$). For our numerical results we use
$\alpha_{s}(m_{t})=0.1073(24)$, which we obtained with the program
{\tt RunDec}~\cite{Chetyrkin:2000yt} from the values
$\alpha_s(m_Z)=0.1176(20)$ and $m_Z =
91.1876(21)$~GeV~\cite{Amsler:2008zzb}. We find
\begin{eqnarray}
  \label{num}
        {\cal F}_{L}&=& 0.6978 - 0.0075 - 0.0023 \nonumber \\
        &=& 0.6978 (1 -0.0108 - 0.0033)\,, \nonumber \\
        {\cal F}_{+}&=& 0 + 0.00103 + 0.00023\,, \nonumber \\
        {\cal F}_{-}&=&  0.3022 + 0.0065 + 0.0021 \nonumber\\
        &=&  0.3022 (1 + 0.0215 + 0.0070)\,.
\end{eqnarray} 
The results in Eq.~(\ref{num}) contain higher orders in
$\alpha_{s}$ from the expansion of the denominators in
Eq.(\ref{helfracdef}). In order to maintain the
constraint ${\cal F}_{L}+{\cal F}_{+}+{\cal F}_{-}=1$, we prefer the
unexpanded definition of helicity fractions (\ref{helfracdef}).
 
The numbers in Eq.~(\ref{num}) show the good convergence of the
perturbative expansion,  even though 
$\Delta{\cal F}_{i}^{(1)}/{\cal  F}_{i}^{(0)}$ (for $i=L,-$) is much
smaller than 
$\Delta{\cal F}_{i}^{(2)}/\Delta{\cal F}_{i}^{(1)}$. The NLO corrections
to the helicity fractions are already close to the expected future
experimental sensitivities and the NNLO corrections increase these by
approximately a third. In particular, the NNLO calculation of the
helicity fraction ${\cal F}_{+}$ remains at the order of $0.001$. 
Should a measurement reveal a significantly larger value, it would be
a clear signal of New Physics. 


\begin{acknowledgments} 
J.G.K.~gratefully acknowledges helpful discussions with S.~Groote and
A.~Kadeer who participated in the early stages of this
investigation. The research of J.H.P.~and A.C.~was supported by Science
and Engineering Research Canada. J.H.P.~was also supported by the
Alberta Ingenuity Foundation and thanks the Graduiertenkolleg
``Eichtheorien'' at the University of Mainz for partial travel support.
Our calculation was done using {\tt FORM}~\cite{Vermaseren:2000nd}. The
Feynman diagrams were drawn with {\tt Axodraw}~\cite{Vermaseren:1994je}
and {\tt JaxoDraw}~\cite{Binosi:2003yf}.
\end{acknowledgments}




\begin{thebibliography}{99}
\bibitem{Affolder:1999mp}
  A.~A.~Affolder {\it et al.}  [CDF Collaboration],
  Phys.\ Rev.\ Lett.\  {\bf 84}, 216 (2000)
  [arXiv:hep-ex/9909042].

\bibitem{Acosta:2004mb}
  D.~E.~Acosta {\it et al.}  [CDF Collaboration],
  Phys.\ Rev.\  D {\bf 71}, 031101 (2005)
  [Erratum-ibid.\  D {\bf 71}, 059901 (2005)]
  [arXiv:hep-ex/0411070].

\bibitem{Abulencia:2005xf}
  A.~Abulencia {\it et al.}  [CDF Collaboration],
  Phys.\ Rev.\  D {\bf 73}, 111103 (2006)
  [arXiv:hep-ex/0511023].

\bibitem{Abulencia:2006iy}
  A.~Abulencia {\it et al.}  [CDF Collaboration],
  Phys.\ Rev.\ Lett.\  {\bf 98}, 072001 (2007)
  [arXiv:hep-ex/0608062].

\bibitem{Abulencia:2006ei}
  A.~Abulencia {\it et al.}  [CDF II Collaboration],
  Phys.\ Rev.\  D {\bf 75}, 052001 (2007)
  [arXiv:hep-ex/0612011].

\bibitem{Aaltonen:2008ei}
  T.~Aaltonen {\it et al.}  [CDF Collaboration],
  Phys.\ Lett.\  B {\bf 674}, 160 (2009)
  [arXiv:0811.0344 [hep-ex]].

\bibitem{Aaltonen:2010ha}
  T.~Aaltonen {\it et al.}  [The CDF Collaboration],
  arXiv:1003.0224 [hep-ex].

\bibitem{Abazov:2005fk}
  V.~M.~Abazov {\it et al.}  [D0 Collaboration],
  Phys.\ Rev.\  D {\bf 72}, 011104 (2005)
  [arXiv:hep-ex/0505031].

\bibitem{Abazov:2004ym}
  V.~M.~Abazov {\it et al.}  [D0 Collaboration],
  Phys.\ Lett.\  B {\bf 617}, 1 (2005)
  [arXiv:hep-ex/0404040].

\bibitem{Abazov:2006hb}
  V.~M.~Abazov {\it et al.}  [D0 Collaboration],
  Phys.\ Rev.\  D {\bf 75}, 031102 (2007)
  [arXiv:hep-ex/0609045].

\bibitem{Abazov:2007ve}
  V.~M.~Abazov {\it et al.}  [D0 Collaboration],
  Phys.\ Rev.\ Lett.\  {\bf 100}, 062004 (2008)
  [arXiv:0711.0032 [hep-ex]].

\bibitem{Kane:1991bg}
  G.~L.~Kane, G.~A.~Ladinsky and C.~P.~Yuan,
  Phys.\ Rev.\  D {\bf 45}, 124 (1992).

\bibitem{Abazov:2009cp}
  V.~M.~Abazov {\it et al.}  [D0 Collaboration],
  Phys.\ Rev.\ Lett.\  {\bf 103}, 141801 (2009)
  [arXiv:0908.0766 [hep-ex]].

\bibitem{CDF2010}
  CDF Collaboration, preliminary results for 2010 Winter Conferences,
  CDF/PHYS/TOP/PUBLIC/10077. 
 
\bibitem{Fujikawa:1993zu}
  K.~Fujikawa and A.~Yamada,
  Phys.\ Rev.\  D {\bf 49}, 5890 (1994).

\bibitem{Cho:1993zb}
  P.~L.~Cho and M.~Misiak,
  Phys.\ Rev.\  D {\bf 49}, 5894 (1994)
  [arXiv:hep-ph/9310332].

\bibitem{AguilarSaavedra:2007rs}
  J.~A.~Aguilar-Saavedra, J.~Carvalho, N.~F.~Castro, A.~Onofre and F.~Veloso,
  Eur.\ Phys.\ J.\  C {\bf 53}, 689 (2008)
  [arXiv:0705.3041 [hep-ph]].

\bibitem{Grzadkowski:2008mf}
  B.~Grzadkowski and M.~Misiak,
  Phys.\ Rev.\  D {\bf 78}, 077501 (2008)
  [arXiv:0802.1413 [hep-ph]].

\bibitem{Fischer:1998gsa}
  M.~Fischer, S.~Groote, J.~G.~K{\"o}rner, M.~C.~Mauser and B.~Lampe,
  Phys.\ Lett.\  B {\bf 451}, 406 (1999)
  [arXiv:hep-ph/9811482].

\bibitem{Fischer:2000kx}
  M.~Fischer, S.~Groote, J.~G.~K{\"o}rner and M.~C.~Mauser,
  Phys.\ Rev.\  D {\bf 63}, 031501(R) (2001)
  [arXiv:hep-ph/0011075].

\bibitem{Fischer:2001gp}
  M.~Fischer, S.~Groote, J.~G.~K{\"o}rner and M.~C.~Mauser,
  Phys.\ Rev.\  D {\bf 65}, 054036 (2002)
  [arXiv:hep-ph/0101322].

\bibitem{Do:2002ky}
  H.~S.~Do, S.~Groote, J.~G.~K{\"o}rner and M.~C.~Mauser,
  Phys.\ Rev.\  D {\bf 67}, 091501(R) (2003)
  [arXiv:hep-ph/0209185].

\bibitem{Wagner:2010wd}
  W.~Wagner  [CDF and D0 Collaboration],
  arXiv:1003.4359 [hep-ex].

\bibitem{Amidei:1996dt}
  D.~Amidei {\it et al.}  [TeV-2000 Study Group],
  Fermilab-Pub-96/046,
  \url{http://www-theory.fnal.gov/TeV2000.html}.

\bibitem{Blokland:2004ye}
  I.~R.~Blokland, A.~Czarnecki, M.~\'{S}lusarczyk and F.~Tkachov,
  Phys.\ Rev.\ Lett.\  {\bf 93}, 062001 (2004)
  [arXiv:hep-ph/0403221].

\bibitem{Blokland:2005vq}
  I.~R.~Blokland, A.~Czarnecki, M.~\'{S}lusarczyk and F.~Tkachov,
  Phys.\ Rev.\  D {\bf 71}, 054004 (2005)
  [Erratum-ibid.\  D {\bf 79}, 019901 (2009)]
  [arXiv:hep-ph/0503039].

\bibitem{Chetyrkin:1999ju}
  K.~G.~Chetyrkin, R.~Harlander, T.~Seidensticker and M.~Steinhauser,
  Phys.\ Rev.\  D {\bf 60}, 114015 (1999)
  [arXiv:hep-ph/9906273].

\bibitem{Smirnov:2002pj}
  V.~A.~Smirnov,
  Springer Tracts Mod.\ Phys.\  {\bf 177}, 1 (2002).

\bibitem{Laporta:1996mq}
  S.~Laporta and E.~Remiddi,
  Phys.\ Lett.\  B {\bf 379}, 283 (1996)
  [arXiv:hep-ph/9602417].

\bibitem{Laporta:2001dd}
  S.~Laporta,
  Int.\ J.\ Mod.\ Phys.\  A {\bf 15}, 5087 (2000)
  [arXiv:hep-ph/0102033].

\bibitem{Pak:2006xf}
  A.~Pak, I.~R.~Blokland and A.~Czarnecki,
  Phys.\ Rev.\  D {\bf 73}, 114009 (2006)
  [arXiv:hep-ph/0604233].

\bibitem{Larin:1993tq}
  S.~A.~Larin,
  Phys.\ Lett.\  B {\bf 303}, 113 (1993)
  [arXiv:hep-ph/9302240].

\bibitem{Larin:1991tj}
  S.~A.~Larin and J.~A.~M.~Vermaseren,
  Phys.\ Lett.\  B {\bf 259}, 345 (1991).

\bibitem{Chetyrkin:2000yt}
  K.~G.~Chetyrkin, J.~H.~K{\"u}hn and M.~Steinhauser,
  Comput.\ Phys.\ Commun.\  {\bf 133}, 43 (2000)
  [arXiv:hep-ph/0004189].

\bibitem{Amsler:2008zzb}
  C.~Amsler {\it et al.}  [Particle Data Group],
  Phys.\ Lett.\  B {\bf 667}, 1 (2008).

\bibitem{Vermaseren:2000nd}
  J.~A.~M.~Vermaseren,
  arXiv:math-ph/0010025.

\bibitem{Vermaseren:1994je}
  J.~A.~M.~Vermaseren,
  Comput.\ Phys.\ Commun.\  {\bf 83}, 45 (1994).

\bibitem{Binosi:2003yf}
  D.~Binosi and L.~Theussl,
  Comput.\ Phys.\ Commun.\  {\bf 161}, 76 (2004)
  [arXiv:hep-ph/0309015].
\end{thebibliography}
\end{document}